**Stabilization of Ferroelectric $Hf_{0.5}Zr_{0.5}O_2$ Epitaxial Films *via* Monolayer Reconstruction Driven by Interfacial Redox Reaction**

Yufan Shen[1], Mitsutaka Haruta[1], I-Ching Lin[1], Lingling Xie[1], Daisuke Kan[1*], Yuichi Shimakawa[1]

[1]Institute for Chemical Research, Kyoto University, Uji, Kyoto 611-0011, Japan

*) dkan@scl.kyoto-u.ac.jp

**Abstract**

The binary fluorite oxide $Hf_{0.5}Zr_{0.5}O_2$ tends to grab a significant amount of notice due to the distinct and superior ferroelectricity found in its metastable phase. Stabilizing the metastable ferroelectric phase and delineating the underlying growth mechanism, however, are still challenging. Recent discoveries of metastable ferroelectric $Hf_{0.5}Zr_{0.5}O_2$ epitaxially grown on structurally dissimilar perovskite oxides have triggered intensive investigations on the ferroelectricity in materials that are nonpolar in bulk form. Nonetheless, the growth mechanism for the unique fluorite/perovskite heterostructures has yet to be fully explored. Here we show that the metastable ferroelectric $Hf_{0.5}Zr_{0.5}O_2$ films can be stabilized even on a one-unit-cell-thick perovskite $La_{0.67}Sr_{0.33}MnO_3$ buffer layer. In collaboration with scanning transmittance electron microscopy (STEM) based characterizations, we show that monolayer reconstruction driven by interfacial redox reactions plays a vital role in the formation of a unique heterointerface between the two structurally dissimilar oxides, providing the template monolayer that facilitates the epitaxial growth of the metastable HZO films. Our findings offer significant insights into the stabilization mechanism of the ferroelectric $Hf_{0.5}Zr_{0.5}O_2$, and this mechanism could be extended for exploring functional metastable phases of various metal oxides.

**Introduction**

Nano-scaled dielectric and ferroelectric materials have shown various remarkable properties, such as giant pyroelectricity, large electrostriction, and energy storage[1–4]. Recently, surging research interests have been drawn towards the ferroelectricity in the meta-stable orthorhombic phase of the fluorite oxide $Hf_{0.5}Zr_{0.5}O_2$ (o-HZO, Fig. 1(A)), whose most stable phase is monoclinic (m-HZO) and non-polar[5–15]. The ferroelectric polarization switching in o-HZO, in contrast to that in conventional ferroelectrics like $Pb(Zr,Ti)O_3$, $BaTiO_3$, and $BiFeO_3$ is demonstrated with significantly suppressed leakage current even in nanometer-thick film specimens, revealing the potential application of HZO for synapse-like computers and nonvolatile memories[16–20]. A key issue in putting ferroelectric HZO into practical application is to stabilize the metastable o-HZO. Although ferroelectricity is seen for metastable o-HZO samples obtained through nonequilibrium growth processes produced by rapid thermal treatments[21,22], undesired nonpolar phases are often unavoidable[23], especially in polycrystalline specimens. Moreover, such mixtures of nonpolar phases deteriorate HZO's ferroelectricity, making it hard to understand the underlying physics behind the ferroelectricity in metastable o-HZO. Alternatively, investigations have revealed that (111)-oriented o-HZO films can be epitaxially stabilized on (001)-oriented perovskite manganite $La_{0.67}Sr_{0.33}MnO_3$ (LSMO) films via heterointerfaces[24,25]. While other perovskite oxides, such as $SrRuO_3$ and $LaNiO_3$, do not enable interfacial stabilization of o-HZO[26]. These observations signify that dissecting LSMO/HZO heterointerfaces is critical for better understanding the stabilization of the metastable polar o-HZO, which would further offer approaches to stabilize metastable phases of other oxides. However, mechanisms for this interfacial stabilization are still highly debated, with several possible routes proposed recently, such as interfacial cation exchanges, interlayer hole transfer from LSMO to HZO, and domain-matching epitaxy[27–29]. The underlying growth mechanism for the unique heterointerface that stabilizes the o-HZO epitaxial films on the perovskite LSMO layers has not been fully clarified.

In this study, we elucidate the phase stability of the metastable o-HZO by tuning the thickness and Sr composition of the LSMO buffer layers on (100) $SrTiO_3$ (STO) substrates, probing the effects that the buffer layers brought on the formations of LSMO/HZO heterointerfaces. We find that the (001)-oriented LSMO down to one unit-cell (u.c.) thickness (0.4 nm) can still stabilize the (111)-oriented o-HZO films, while phase-mixed and amorphous HZO films were grown on $LaMnO_3$ (LMO) buffer layers and on STO substrates, respectively. Combining this finding with the results of characterizations by high-angle annular dark-field (HAADF) imaging, energy-dispersive-x-ray spectroscopy (EDS), and electron energy-loss spectroscopy (EELS) in scanning transmission electron microscope (STEM), we argue that monolayer reconstructions assisted by interfacial redox reactions between HZO and LSMO promote the formation of the unique heterointerface and lead to the stabilization of the metastable epitaxial o-HZO films.

**Results**
**Structural phases and ferroelectricity properties of HZO thin films**

We employed pulsed laser deposition and deposited HZO films on $TiO_2$-terminated (001) STO substrates buffered by LSMO films whose thicknesses ranged from 0 to 45nm, and by 45-nm-thick LMO films. The x-ray 2θ/θ diffraction patterns for the grown heterostructures are shown in Fig. 1(B). No peaks except the ones from the substrates were found when HZO films were deposited directly on STO substrates, signifying the amorphous growth of HZO thin films. On the other hand, inserting LSMO and LMO epitaxial buffer layers is found to trigger and strongly influence the epitaxial growth of HZO films. HZO films deposited on (001)-oriented LSMO buffer layers exhibit the (111) reflections at 2θ ~30° together with thickness fringes. No reflections from any HZO phases other than o-HZO are observed. These results indicate that the metastable o-HZO films with the (111) orientation can be stabilized on the LSMO buffer layers, which agrees with previous reports[23,30,31]. Omega scan results show that the full width at the half maximum of the (111) o-HZO reflection is as narrow as 0.06°, confirming the good crystallinity of our stabilized o-HZO films (Fig. S1). Interestingly, this LSMO-layer-assisted stabilization and epitaxial growth of o-HZO films are independent of the LSMO layers' thickness. As shown in the figure, even the monolayer-thick (0.4 nm) LSMO layer allows the epitaxial growth of (111)-oriented o-HZO films without any secondary phases. In contrast, inserting LMO layers (0% Sr concentration in LSMO) results in the growth of a mixture consisting of both (111)-oriented o-HZO films and polycrystalline monoclinic HZO films. We note that both LSMO and LMO buffer layers are of good crystallinity and have surfaces as flat as those of STO substrates, which are ensured by the observations of clear thickness fringes around the (001) fundamental reflection in the 2θ/θ diffraction patterns and by atomic force microscope (AFM) topography (Fig. S2). These results rule out that the formation of polycrystalline HZO films stems from the surface roughness of the buffer layers, implying that some other effects at the LSMO/HZO interface play a critical role in the stabilization and epitaxial growth of the o-HZO films. We further confirmed the ferroelectricity of the o-HZO films stabilized on the LSMO buffer layers. Fig. 1(C) shows the room-temperature polarization-electric field (P-E) and current-electric field (I-E) loops for the LSMO(45)/HZO heterostructures. Both P-E and I-E loops show clear hysteresis associated with electric-field-induced polarization switching. A remnant ferroelectric polarization of the o-HZO films is as large as 12 μC/cm$^2$, which is comparable to those seen in previous reports[27,32–34].

**HAADF-STEM characterization for the HZO/L(S)MO heterointerfaces**

We carried out cross-sectional STEM observations to examine interfaces between HZO films and the buffer layers. Fig. 2, (A–D) shows HAADF-STEM images, with the zone axes along the [010] directions of the substrates, for the LMO(45)/HZO, LSMO(45)/HZO, LSMO(0.8)/HZO, and LSMO(0.4)/HZO heterostructures respectively. As expected from the results of x-ray diffraction characterizations (Fig. 1(B)), the image contrasts seen in the HZO films on LSMO (0.4, 0.8, and 45 nm-thick) buffer layers correspond to the (111)-oriented orthorhombic phase of HZO[25,28,35]. Because (111)-oriented crystallographic domains with different in-plane orientations exist along the zone axes, the Hf/Zr atomic columns in the o-HZO films are not well resolved along the horizontal direction (Fig. S3). On the other hand, in the HZO layer deposited on the LMO buffer layer, the different image contrast originating from the (001)-oriented monoclinic phase of HZO is seen, in agreement with the observation that the LMO buffer layer leads to the growth of a mixture of o-HZO and m-HZO films (Fig. 1(B)). Interestingly, the structural phases of the HZO films are also found to depend on interface structures between the HZO films and the buffer layers, indicating that the L(S)MO/HZO interfaces play critical roles in the phase stabilization and epitaxial growth of the HZO films. As shown in Fig. 2(E–H), the interfaces between the o-HZO films and the LSMO layers (marked in blue) are sharp, regardless of the thickness of the LSMO layer. Given that the LSMO layers are grown on the $TiO_2$-terminated (the B-site-layer-terminated) STO substrates, the $MnO_2$ layer should be the topmost layer of the LSMO layer and forms an interface with the HZO films[36,37]. However, in the LSMO/o-HZO interface region, no image contrasts corresponding to the Mn atomic columns are seen. Furthermore, the LSMO layer in the LSMO(0.8)/o-HZO heterostructure is approximately 0.6 nm (1.5 unit cells)-thick (marked by purple box), seemingly terminated by the (La,Sr)O layer. The LSMO/o-HZO interfaces are found to consist of atomic columns whose image contrasts are much brighter than those of Mn atoms. These observations imply that the interfaces are reconstructed when the HZO films are deposited on the LSMO layer. In addition, when the LSMO thickness is reduced to 0.4 nm (one unit cell thick), an interfacial monolayer whose image contrasts are similar to those for the LSMO(0.8)/o-HZO interface is still observed (Fig. 2(H)), signifying that structural reconstruction at the LSMO/HZO interfaces associates with only the topmost $MnO_2$ layer of the LSMO buffer and plays an important role on the stabilization of metastable o-HZO. We also note that although some structural defects related to cation off-stoichiometry and oxygen vacancies might be accommodated into LSMO layers through variations in the growth conditions such as laser fluence and oxygen pressures, LSMO layers deposited with growth conditions deviating from the optimal condition can still stabilize o-HZO films on top of them (Fig. S4), implying that such structural defects have little effect on the o-HZO stabilization. Furthermore, the LMO/HZO interface (Figs. 2(A), 2(E) and S5) is found not to be as sharp as the LSMO/HZO interface, and the m-HZO films are grown on the disordered interface, while the o-HZO films are grown on the well-ordered interface. Our observations indicate that the interfacial-reconstruction-induced monolayer led to the stabilization and epitaxial growth of o-HZO.

**Elemental mapping for the LSMO/HZO heterointerfaces**

EDS elemental mapping further reveals the structural reconstruction in the LSMO/HZO interfaces. Fig. 3 shows the results for the heterointerfaces of LSMO(45)/o-HZO, LSMO(0.8)/o-HZO, and LSMO(0.4)/o-HZO. For the LSMO(45)/o-HZO case (Fig. 3(A)), the LSMO regions showed EDS signals of La, Sr and Mn, whose positions agree well with those expected from the perovskite structure of LSMO. Furthermore, the detections of La signals from the LSMO's topmost layer tell that the underneath LSMO(45) buffer is indeed (La,Sr)O-terminated rather than $MnO_2$-terminated as would have been expected from the substrates' surface termination. We also found that the Hf and Zr signals are detected only in the HZO layer region above the topmost (La,Sr)O layer. These observations indicate that the bright HAADF image contrasts seen on the (La,Sr)O topmost layer are mainly contributed by Hf and Zr (Fig. 2). For the LSMO(0.8)/HZO heterostructure, as shown in Fig. 3(B), the STO substrates maintain the $TiO_2$-layer termination, while the LSMO region consists of one $MnO_2$ layer sandwiched by two (La,Sr)O layers, showing that the 0.8-nm-thick LSMO layer is terminated with the (La,Sr)O-layer, as observed for the LSMO(45)/HZO interface. The Hf and Zr EDS signals are detected in the region above the topmost (La,Sr)O layer as well, and they become apparently weaker in this interfacial HZO layer, whose HAADF-STEM image contrasts are seen in the same positions as those of the Mn atom columns in the LSMO layer. The elemental mapping for the heterostructure with the 0.4 nm (one unit cell)-thick LSMO layer further corroborates the scenario that the monolayer on the topmost (La,Sr)O layer is $(Hf,Zr)O_x$, formed through structural reconstruction (or interfacial reactions) between LSMO and HZO. As displayed in Fig. 3(C), this interfacial $(Hf,Zr)O_x$ monolayer even forms on the (La,Sr)O monolayer located on the topmost $TiO_2$ layer of the substrate, signifying that Hf and Zr atoms deposited on the one-unit-cell-thick LSMO layer substitute Mn atoms within its $MnO_2$ plane, forming the interfacial $(Hf,Zr)O_x$ monolayer and promoting the epitaxial growth of o-HZO.

**Intralayer charge transfer at the interfacial $(Hf,Zr)O_x$ monolayer**

Fig. 4 shows the spatially resolved EELS spectra of the Mn $L_{2,3}$-edge across the LSMO/o-HZO heterointerfaces. Surprisingly, the energy loss of the Mn $L_{2,3}$-edges can be detected even in the interfacial $(Hr,Zr)O_x$ monolayers in all the LSMO/o-HZO heterostructures, indicating that some Mn atoms reside in the $(Hr,Zr)O_x$ monolayer after the Mn-to-Hf/Zr exchange. Furthermore, the energy loss of the Mn $L_3$-edge at the $(Hr,Zr)O_x$ monolayer is lowered by ~1.7eV than those in the LSMO(45) and LSMO(0.8) layers, revealing the intralayer electron transfer to Mn and the lowering of the Mn valence state at the $(Hr,Zr)O_x$ monolayer[38]. Interestingly, the Mn $L_3$-edge energy loss lowering at the $(Hr,Zr)O_x$ monolayer in the LSMO(0.4)/o-HZO heterostructure is almost the same as those for the LSMO(45)/o-HZO and LSMO(0.8)/o-HZO cases, further demonstrating that the intralayer electron transfer is associated with the Mn-to-Hf/Zr exchange. The electron transfer to Mn within the $(Hr,Zr)O_x$ monolayer in the LSMO(0.4)/o-HZO heterostructure rules out that interlayer electron transfers from the LSMO buffer to the HZO layer are responsible for the observed lowering of the Mn energy loss. Moreover, it is unlikely that ferroelectric polarizations in o-HZO films result in the interfacial lowering of the Mn valence state because polarization-induced effects usually spread over 3-4 unit cells-thick regions with respect to the interface[39,40]. Our observations indicate that the interfacial reactions triggered by the Mn-to-Hf/Zr exchange and intralayer electron transfer play a key role in forming the unique monolayer-thick HZO/LSMO heterointerface and stabilizing (111)-oriented o-HZO epitaxial films.

**Discussion**

We now discuss how the unique interface forms between the o-HZO epitaxial films and the LSMO buffer layers. Our STEM-based characterizations show that the (Hf,Zr)O$_x$ (x < 2) monolayer bridges these two structurally different oxides, and this monolayer forms on the topmost (La,Sr)O layer, not on the MnO$_2$ layer of the LSMO layer. One thing that should be noted is that depositing HZO films directly on both (La,Sr)O-terminated LSMO layers and LaO-terminated LMO layers leads to the formation of a mixture of o- and m-HZO films (Fig. S6), which agrees with the previous report[27]. These observations highlight that simply depositing and oxidizing Hf/Zr atoms on the topmost (La,Sr)O layer leads to the most-stable m-HZO, and that the interfacial Mn-to-Hf/Zr exchange in the MnO$_2$-terminated layers of the LSMO buffer is crucial for stabilizing the metastable o-HZO through the forming (Hf,Zr)O$_x$ monolayer. Recent theoretical calculations[28] also support the interface reconstruction scenario, as they showed that when Hf atoms are deposited on a MnO$_2$-terminated LSMO layer under oxygen-deprived conditions, they energetically prefer to replace the Mn atoms in the topmost MnO$_2$ layer. It should be underlined that this Mn to Hf/Zr exchange process can be regarded as a redox reaction between the topmost MnO$_2$ layer and Hf/Zr adatoms, providing a route for oxidizing Hf and Zr adatoms without forming the most stable m-HZO phase. Given the low ionization energies of Hf and Zr, both elements tend to have the stable valence state of 4+. However, under oxygen-deprived (relatively reducing) conditions in which the cation exchange and o-HZO stabilization are preferred[41,42], oxidizing Hf and Zr to valence states close to 4+ without forming m-HZO is rather difficult. Therefore, the interfacial redox reaction associated with the Mn to Hf/Zr exchange and the intralayer electron transfer enable stabilizing $Hf^{4+}$ and $Zr^{4+}$ at the LSMO/HZO interface, which leads to the formation of the $(Hf,Zr)O_x$ monolayer and the epitaxial growth of the metastable o-HZO. It is worthy of addressing that the ability of Mn to accommodate electrons related to the intralayer charge transfer depends on the Sr composition (or the Mn valence state) in the LSMO layers. Mn with higher valence states would prefer accommodating electrons from Hf/Zr and driving the interfacial redox reaction. For the LMO layers, the topmost $MnO_2$ layer consists of $Mn^{3+}$ only, which does not easily accommodate extra electrons from the Hf/Zr atoms. This can explain why the LMO/HZO interface remains disordered with both m- and o-HZO films grown on top of it (Fig. 2).

**Summary**

In summary, we identified the unique $(Hf,Zr)O_x$ monolayer that bridges the structurally dissimilar fluorite o-HZO and perovskite LSMO and showed that the formation of the interfacial monolayer on the (La,Sr)O topmost layer of the LSMO buffer is independent of the LSMO layers' thickness. Our observations indicate that the redox reactions between the $MnO_2$ termination layer and Hf/Zr adatoms lead to the formation of the $(Hf,Zr)O_x$ monolayer through the cation exchange assisted by the intralayer charge transfer, which facilitates the epitaxial growth of the metastable o-HZO films. Our results offer critical insight into the stabilization mechanism of metastable ferroelectric o-HZO, which can be extended to explore functional metastable phases of metal oxides [43–47].

**Materials and Methods**

**Sample Fabrication**

All samples were fabricated on TiO$_2$-terminated (001) SrTiO$_3$ (STO) substrates (Shinkosya Co., Japan) by using pulsed laser deposition (PLD) with a KrF excimer laser ($\lambda$ = 248 nm). The La$_{0.67}$Sr$_{0.33}$MnO$_3$ (LSMO) and LaMnO$_3$ (LMO) buffer layers were deposited at 650 °C and under the oxygen partial pressures of 100 and 10 mTorr, respectively, to obtain the optimal crystallinity for each buffer layer. During the deposition, the LSMO and LMO ceramic targets were ablated with a laser fluence of 1.2 J/cm$^2$ and at the repetition frequency of 5 Hz. The Hf$_{0.5}$Zr$_{0.5}$O$_2$ (HZO) thin films were subsequently deposited on the buffer layers without breaking the vacuum of the PLD chamber. The HZO depositions were made at 800 °C and under the oxygen partial pressure of 75 mTorr. The HZO ceramic targets were pulsed with a laser fluence of 1.4 J/cm$^2$ at 2 Hz. After the depositions, the samples were cooled down to room temperature with a partial oxygen pressure of 75mTorr.

**Structural Characterizations**

X-ray 2θ/θ diffraction measurements were performed with a lab-source four-circle diffractometer (X'Pert MRD, PANalytical) using the Cu K$_{\alpha 1}$ radiation. Cross-sectional STEM samples were prepared by a focused ion beam system (JIB-4700F). STEM experiments were conducted at room temperature on a spherical aberration-corrected STEM (JEM-ARM200F) equipped with an EDS spectrometer (JED-2300T) and an EELS spectrometer (Gatan Quantum ERS). The experiments were performed at 200 kV. The probe convergence semi-angle was 24.6 mrad and the EELS collection semi-angle was 57.3 mrad. EELS measurements were carried out using dual EELS mode to discuss the correct energy shift, where energy dispersion was 0.1 eV/pixel. The EEL spectra were acquired for a long time by scanning the electron probe along equivalent columns along the in-plane to confirm the position of the electron beam and to improve the signal-to-noise (S/N) ratio.

**Ferroelectric polarization Characterizations**

Polarization hysteresis loops were measured using a ferroelectric analyzer (FCE-10 series TOYO TECH) at room temperature. The top electrodes of 40nm-thick Au pads, which were photolithographically patterned with a diameter of 30 μm were thermally evaporated at room temperature.

**Acknowledgments**

This work was partly supported by Grants-in-Aid for Scientific Research (Nos. 19H05816, 19H05823, 21H01810, 22KK0075, and 23H05457) and by grants for the Integrated Research Consortium on Chemical Sciences and the International Collaborative Research Program of the Institute for Chemical Research in Kyoto University from the Ministry of Education, Culture, Sports, Science, and Technology (MEXT) of Japan. The work was also supported by Japan Science and TechnologyAgency (JST) as part of the Advanced International Collaborative Research Program (AdCORP), Grant Number JPMJKB 2304.

**Figures**

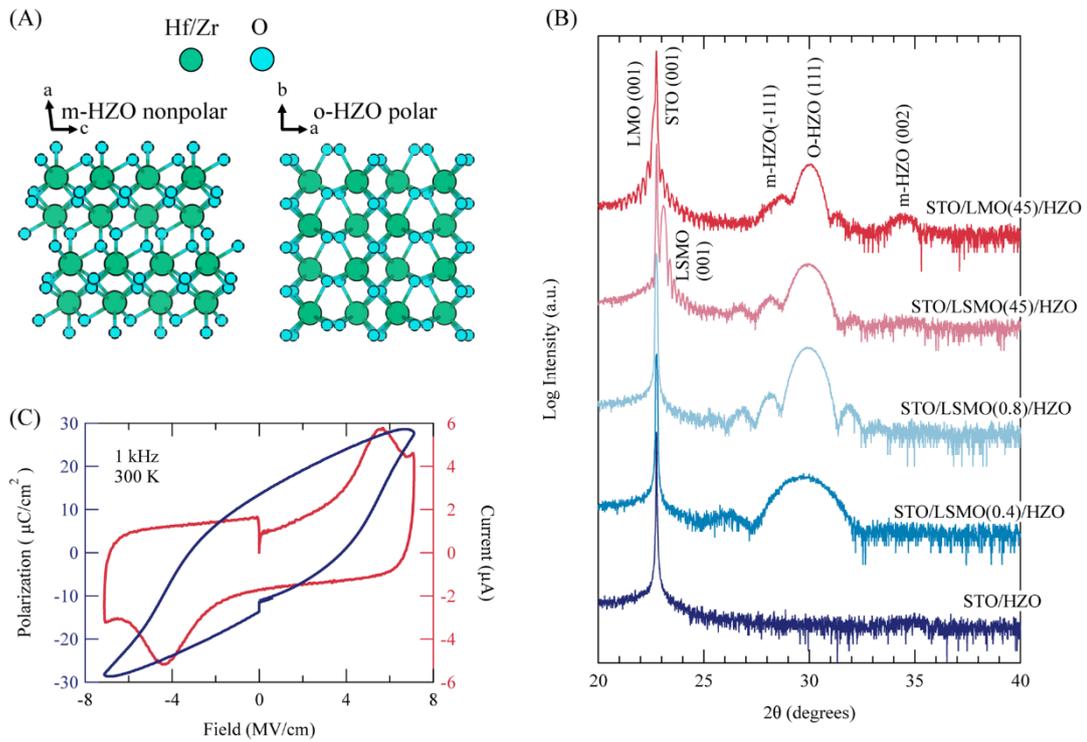

**Fig. 1. Structural phases and ferroelectric properties of Hf$_{0.5}$Zr$_{0.5}$O$_2$ (HZO) thin films. (A)** Crystal structures of the most stable and non-polar monoclinic phase (left; m-HZO), and polar orthorhombic phase (right; o-HZO) of HZO. **(B)** X-ray 2θ/θ diffraction patterns for HZO films deposited on various buffer layers. The numbers in brackets denote the thickness of the L(S)MO buffer layers of the fabricated heterostructures in the unit of a nanometer. The thickness of the HZO layer in the heterostructure is 8 nm, except that the one in the LSMO(0.4)/HZO heterostructure is 4 nm thick. **(C)** Room temperature P-E and I-E hysteresis loops for the LSMO(45)/HZO heterostructure. We note that the P-E loop measurements were performed on more than ten different capacitor samples and confirmed the reproducibility of the ferroelectric hysteresis loops for the o-HZO films.

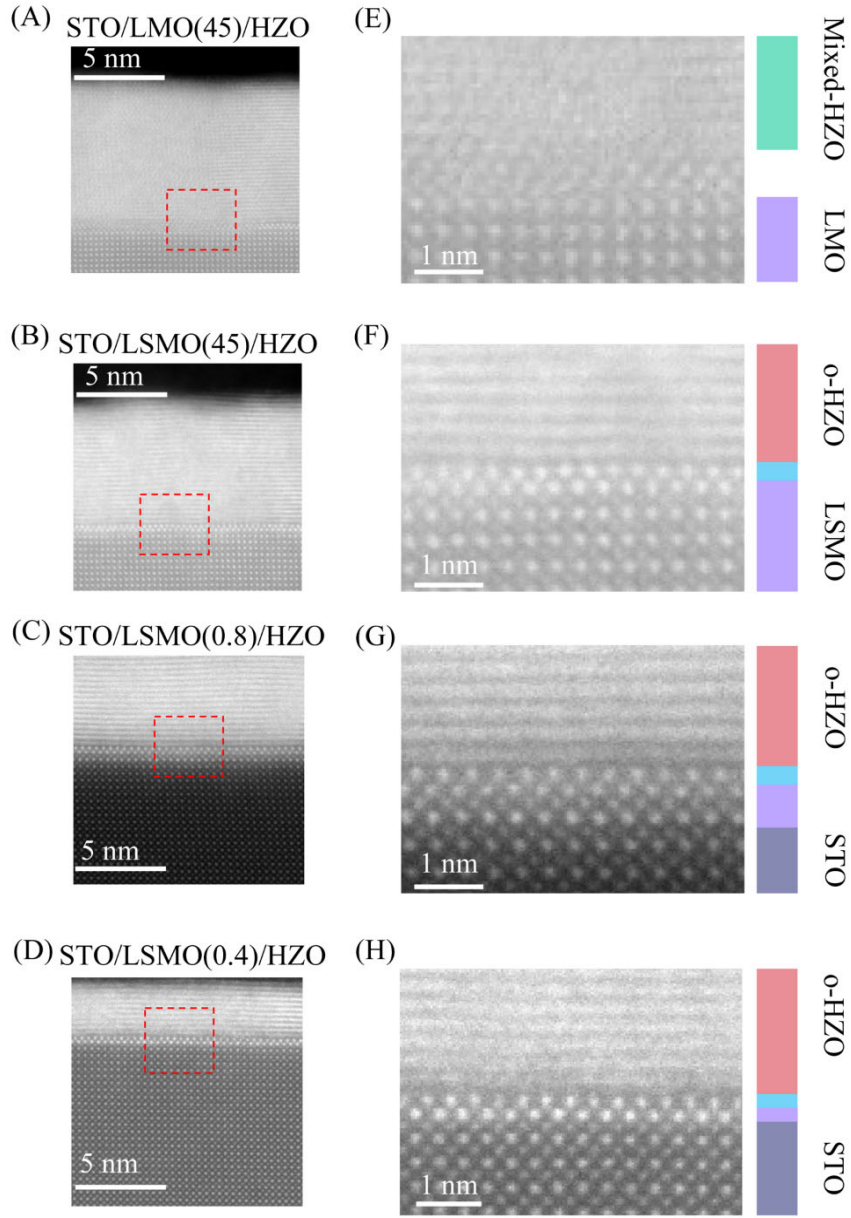

**Fig. 2. Interface structures between HZO and L(S)MO.** **(A-D)** Cross-sectional HAADF-STEM images for the **(A)** LMO(45)/HZO, **(B)** LSMO(45)/HZO, **(C)** LSMO(0.8)/HZO and **(D)** LSMO(0.4)/HZO heterostructures. The enlarged version of 2(a) is provided as Fig. S5 in the supplementary information. **(E-H)** Zoomed images taken from the area framed in red in the corresponding left images. In the figures, the mixed

HZO (a mixture of m- and o-HZO), o-HZO, interfacial monolayer, L(S)MO buffer layer, and STO regions are colored in green, red, blue, purple, and navy, respectively.

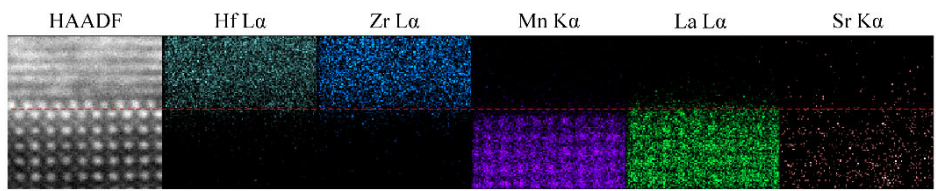

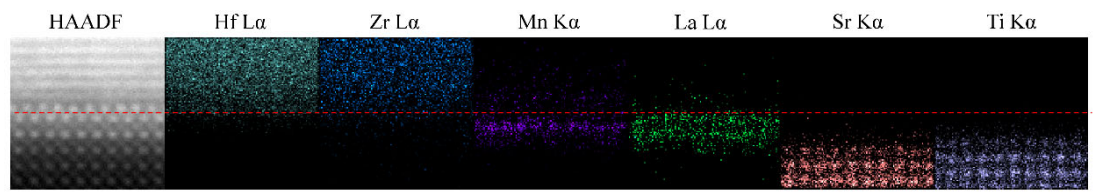

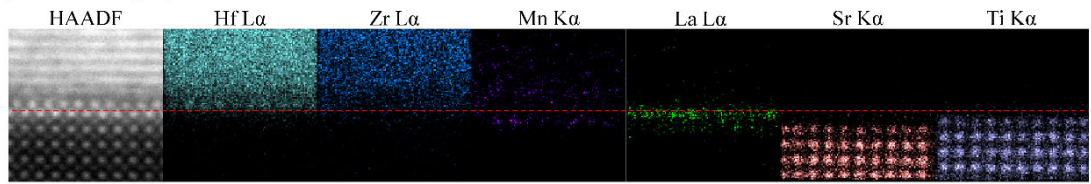

**Fig. 3. Interfacial reconstruction between HZO and LSMO.** Atomically resolved STEM-EDS elemental mapping for the **(A)** LSMO(45)/HZO, **(B)** LSMO(0.8)/ HZO, and **(C)** LSMO(0.4)/HZO heterostructures.

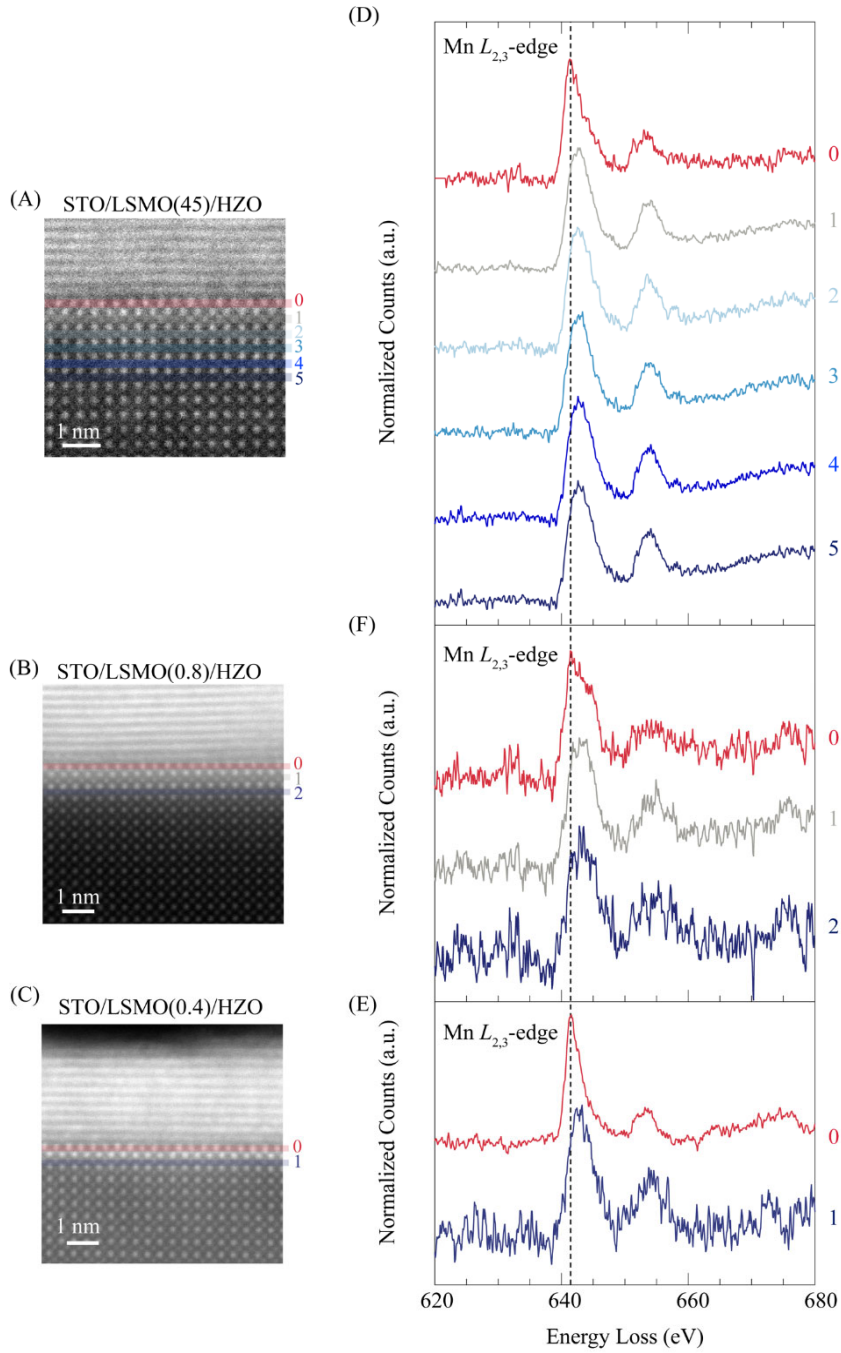

**Fig. 4. Intralayer charge transfer in the interfacial (Hf,Zr)O$_x$ monolayer. (A-C)** HAADF-STEM image for (A) LSMO(45)/o-HZO, (B)LSMO(0.8)/o-HZO, and (C)LSMO(0.4)/o-HZO heterointerfaces. In the images, the interfacial (Hf,Zr)O$_x$ monolayer is numbered zero and colored in red. **(D-F)** Spatial dependence of EELS

spectra of the Mn $L_{2,3}$-edges across the heterointerfaces. Each spectrum was taken at the corresponding numbered atomic positions in the HAADF-STEM image. The measured spectra are normalized in such a way that the Mn $L_3$-edge peak counts become unity.